\documentclass[pre,aps,twocolumn]{revtex4}
\usepackage{graphicx,mathtools,morefloats}

\begin{document}

\title{Quenching to fix metastable states in models of 
prebiotic chemistry}

\author{Qianyi Sheng, Ben Intoy  and J. W. Halley}
\address{School of Physics and
Astronomy, University of Minnesota, Minneapolis, MN 55455}

\date{\today}

\begin{abstract}
For prebiotic chemistry to succeed in producing a starting 
metastable, autocatalytic and reproducing system subject to
evolutionary selection it must satisfy  at least two apparently
contradictory requirements: Because such systems are rare,
a search among vast numbers of molecular combinations must 
take place naturally, requiring rapid rearrangement and breaking of 
covalent bonds.  But once a relevant system is found, such rapid
disruption and rearrangement would be very likely to destroy
the system  before much evolution could take place.  In this paper we 
explore the possibility, using a model developed previously,
that the search process could occur under different environmental conditions
than the subsequent fixation and growth of a lifelike chemical
system. We use the example of a rapid change in temperature
to illustrate the effect and refer to the rapid change as a 
`quench' borrowing terminology from study of the physics and
chemistry of glass formation. The model study shows that 
interrupting a high temperature  nonequilibrium state with a rapid 
quench to lower temperatures can substantially increase the
probability of producing a chemical state with lifelike
characteristics of nonequilibrium metastability, internal
dynamics and exponential population growth in time.  Previously
published data on the length distributions of proteomes of
prokaryotes may be consistent with such an idea and 
suggest a prebiotic  high temperature `search' phase near the
boiling point of water. A rapid change in pH could have 
a similar effect.  We discuss possible scenarios on
early earth which might have allowed frequent quenches of 
the sort considered here to have occurred. The models show 
a strong dependence of the effect on the number of chemical
monomers  available for bond formation. 
\end{abstract}  

\pacs{PACS numbers: }

\maketitle

\section{Introduction} \label{sec:Introduction}

Estimates of the likelihood of natural formation of an initial genome in 
`genome first' models of prebiotic evolution exhibit such small numbers that
the production of such a starter genome by natural  nonbiotic processes 
appears to be nearly impossible (`Eigen's  paradox')\cite{eigen}.  Statistical estimates
of the likelihood of formation of random prion or amyloid-like combinations of
amino acids\cite{mauryprions,reisnerprions,baskakovprions,portilloprions}
 are presumably somewhat higher, though quantitative estimates
do not appear to be available. However even the latter scenario would face the
problem that, in an environment in which many combinations of amino acids
form and then deteriorate, (a `dynamical chemical network'\cite{taran}) it appears 
quite likely that a promising combination would deteriorate before it could
be fixed and begin to grow and reproduce. 

Here we explore the possibility that rapid `quenches' of a dynamical chemical
network (possibly of polypeptides, though our models are not chemically
specific) either by rapid temperature reduction or by change in pH,
might stabilize systems with lifelike characteristics, thereby increasing
the probability of their formation and growth. Such rapid quenches might occur,
for example when material is rapidly ejected from an ocean trench, though other
scenarios can be envisioned.Experiments  based
on that  idea have been reported\cite{matsunoa}-\cite{matsunoc}
 and
  did demonstrate that quenching results in enhanced polypeptide formation.

\begin{figure}
\includegraphics[angle=0,width=3.5in]{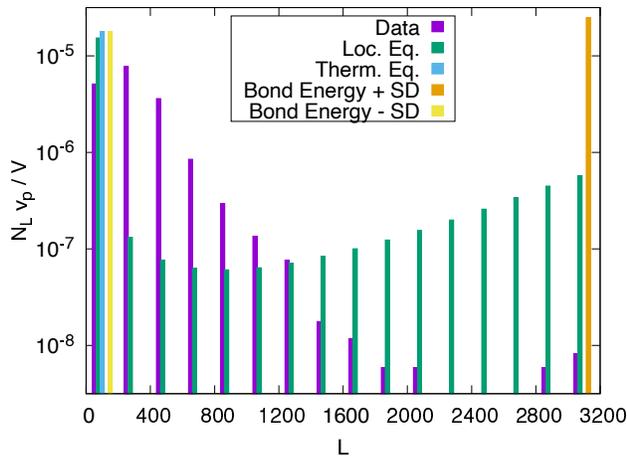}
\caption{\label{fig:MartStdDev}
Data on prokaryote cva  (labeled `Data', purple bars)compared to equilibria at different
values of $\beta \Delta$.  Green bars labelled `local equilibrium' show 
the equilibrium arising from  the local temperature derived from the energy $E$ 
and the particle density $N/V$ giving $\beta \Delta=-2.99$ for this case. 
Blue bar at left labelled `Thermal Equilibrium' shows the equilibrium associated with
ambient external temperature of 293K corresponding to $\beta \Delta=-3.78$.  
Orange bar labelled `Bond energy $+$ SD' shows the population distribution
which would arise at ambient external temperature if the value 
of $\Delta$ were shifted from its average observed value to that value
plus the observed standard deviation of the empirical distribution of bond energies
giving $\beta \Delta=-2.28$. (The orange bar is not visible because it is identical
to the blue bar.)  `Bond energy $-$ SD' (yellow bar) shows the population distribution
which would arise at ambient external temperature if the value 
of $\Delta$ were shifted from its average observed value to that value
minus the observed standard deviation of the empirical distribution of bond energies
giving $\beta \Delta=-5.28$. Note the dramatic sensitivity of the 
equilibrium distribution corresponding to ambient temperature arising from 
varying the bond energy over its range of uncertainty and the similarity
of the observed distribution to an equilibrium distribution arising from the
much higher `local' temperature (green bars) derived from the energy and the particle 
density. 
The values  $-2.2$ kcal/mol and $0.875$ kcal/mol for the average
and standard deviation respectively of the protein bond energies 
are from \cite{martin1998free}. Figure is adapted from \cite{intoyhalleypopulations}. 
}\end{figure}

Before we became aware of references \cite{matsunoa}-\cite{matsunoc}, the idea 
of a rapid quench as a generator of lifelike 
systems was suggested independently to us by our previous 
studies of 
the statistical distribution of polymer lengths in the proteomes of
4,555 prokaryotes\cite{intoyhalleypopulations}.  Some representative data from that paper
is shown in the Fig. \ref{fig:MartStdDev}.

\begin{figure}
\includegraphics{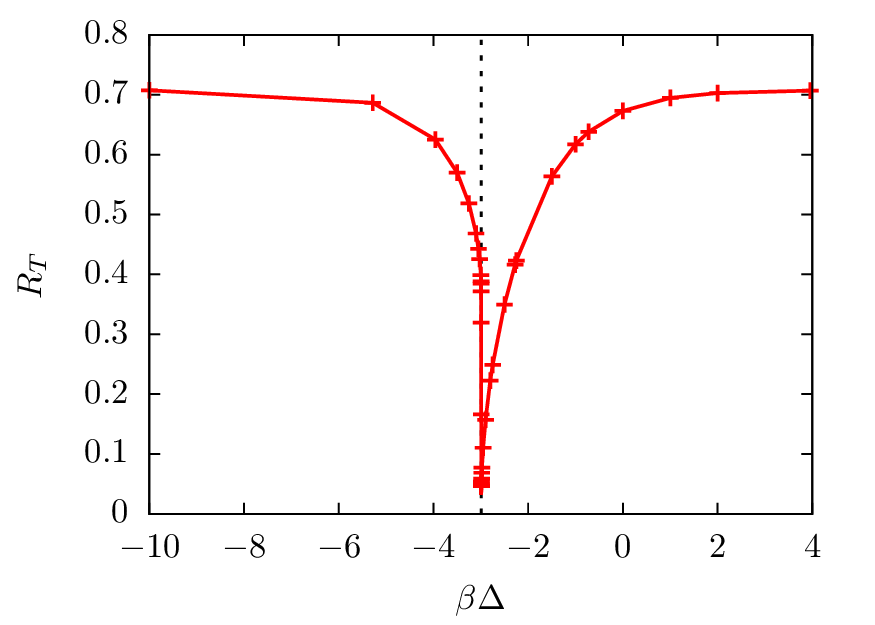}
\caption{\label{fig:RTfuncdb}
  $R_T$ as a function of $\beta \Delta$ for the prokaryote 
  prokary-
  ote Corynebacterium variable DSM 44702 (KEGG code cva
  From Fig. 5 of reference \cite{intoyhalleypopulations}. Here
$R_T$, defined in equation \ref{eq:RG}, is a measure of disequilibrium,
$\beta=1/k_BT$ where $T$ is the absolute temperature and $\Delta$  is
the binding energy of a monomer-monomer bond (peptide bond in this
application). The dashed vertical line is at $\beta\Delta =-\ln 20$, 
the predicted value of the minimum arising from the hypotheses tested
in this paper.
  }
  \end{figure}

The key point is that the length distribution  varies quite smoothly across 
the range of lengths up to about 2200 monomers , 
whereas, if we use measured peptide 
bond energies and room temperature we find equilibrium distributions corresponding
to essentially all very long  polymers (the yellow bar in the figure), or, 
with a slight adjustment of
the bond energy parameter, all dissociated amino acids (the blue bar). It is hard to see 
how
the length distribution observed could have ever been close to an equilibrium
one at room temperature. If the ambient temperature were higher,
then the length distribution could be more uniform (as shown,for example,
in the green bars in
Fig.1) , but in such an environment
peptide bonds would be continually breaking and reforming in a dynamical chemical
network\cite{taran}.  Using the proteome  population data in another way on the
same system we calculated the quantity
\begin{equation}
R_T=\sqrt{\sum_{L} (N_{L}- \overline{N_L(\beta\Delta)}))^2}/(\sqrt{2}V\rho)
\label{eq:RG}
\end{equation}
as a function of $\beta\Delta$ as a measure of how far from equilibrium
the observed polymer length distribution $\{N_L\}$ was from equilibrium at
a temperature $T=1/k_B\beta$ .  Here $\rho$ is the volume density of polymers 
and $V$ is the system volume. With $l_{max}$ defined as the maximum
observed polymer length, $R_T$ is a Euclidean distance in the
$l_{max}$ dimensional space of the $l_{max}$ tuples $\{N_L\}$ normalised to
lie between 0 and 1. $\overline{N_L(\Delta\beta)}$ is the equilibrium
distribution when the system is exposed to a
thermal bath with an external temperature $1/k_B\beta$. $\Delta$ is the
binding energy of the covalent bonds connecting the monomers of the
model.  For peptides it is negative, and we treat mainly that case here.
 
The result from that paper 
, shown in  Fig. \ref{fig:RTfuncdb} indicates that the distribution observed 
would be close to equilibrium (as indicated by the small value of $R_T$, 
) with a thermal bath at a temperature around 400K (using the value 
-2.2 kcal/mol , taken from reference \cite{martin1998free}), 
for the peptide bond energy, negative because the bond is unstable in
aqueous solution. We suggested that this might be 
an indication that the precursor to the proteome formed 
at that high temperature, and then became fixed by rapid quenching to
a lower temperature closer to current ambient temperature as anticipated
in the experiments reported in\cite{matsunoa}-\cite{matsunoc}.

We also noted that the evaluation of a so-called `local' measure $R_L$
, defined in detail in the next section (equation (\ref{eq:RL})), 
of the distance of the observed length distribution from isolated equilibrium
, (see also \cite{intoyhalleypopulations}) and
also obtained from the proteome data, 
was close to the minimum value found from the data displayed in
Fig.  \ref{fig:RTfuncdb}. ($R_L$ is defined by an equation identical to
(\ref{eq:RG}) except that the equilibrium distribution is determined by
use of the total energy of the polymer system as well as the total polymer
density and the temperature $1/k_B\beta$ is not fixed, but is determined
by maximization of the entropy, given the energy and the density.  Please
see
\cite{intoyhalleypopulations} and the next section for more details.)
That suggested to us that the precursor had
formed in an isolated, nearly equilibrium system at around 400K and the 
population distribution had
then been fixed by quenching. Finally we noted that the value of 
400K, given the value of the average peptide bond energy $\Delta$ reported in
\cite{martin1998free} gave a value
of $\beta\Delta$ quite close (vertical dashed line in the
figure) to $-\ln b$ when $b$, the number of available monomers, takes the 
value $b=20$, the well-known 
number of amino acids used in forming the proteins of the biosphere.
In the so-called `Gibbs limit' of our equilibrium expressions for
equilibrium length distributions, as defined in \cite{intoyhalleypopulations} and also in 
the Appendix to  
this paper, that value of $\beta\Delta$ would
give an average value  of $\delta \overline{N_L}/\delta L$ of zero,  thus allowing 
a wide and rapid exploration of the polymer state space at that 
temperature.  (Here $\delta \overline{N_L}$ and $\delta L$ denote finite increments.)So our third suggestion was that quenches from that
temperature were more likely to lead to formation of a rare,
autocatalytic system with lifelike properties, such as disequilibrium, metastability,
high rates of reaction and exponential population growth, after quench because
more possibilities were explored in high temperature states at 
that temperature. Finally we suggested that, at large $b$, the
temperature at which the average value of $\delta \overline{N_L}/\delta L$ is zero would be such that
$\beta\Delta \approx -\ln b$.

In this paper, we report model simulation results which test these
ideas by explicitly modeling the conjectured quenches with
our  Kauffman-like model previously reported in reference
\cite{conditions}. The model is an abstract, or one may say
coarse-grained, description of the real proteome system.  The only
entropic effects which are included are those associated with
the combinatorial degeneracies associated with the availability
of a more than 1 monomer in polymer formation. That abstraction
permits us, in the model, to focus attention on the effects of
such degeneracies, which are the root cause of Eigen's paradox
as discussed above.  Since these combinatorial, that is informational,
 aspects of the
entropy are associated with the formation and dissociation
of covalent bonds which are significantly stronger than the
hydrogen bonds and van der waals forces which determine more
detailed aspects of biopolymer chemistry, we say that our model
is coarse grained in energy, implicitly averaging over such smaller
energy effects (which of course are essential for more detailed modeling
of life). Our simulation
resources limit us to computations up to 7 ($=b$) types of monomers.
(Some statements about the large $b$ limit will turn out to be possible
by analytical means.) However, as we will discuss in detail in section
4, several aspects of the above conjectures turn out to be  consistent with
 results
from the simulated quenches.

In the next section we describe the  model of
reference \cite{conditions} and indicate how quenches were simulated and what 
was measured in the simulation data.
The third section 
states our four conjectures concerning the expected results of the 
simulations based on the qualitative discussion above.  Section 4 
summarizes results of the simulations compared with those expectations.
Section 5 contains conclusions and a brief discussion of the conditions
in ocean trenches and ridges which might result in huge 
numbers of the envisioned quenches occurring over millions of years and
resulting in a substantial probability of fixing a rare, autocatalytic
network with lifelike properties.

\section{Model and Simulation Methods} \label{sec:Simulations}

The model used for quench simulations is fully described in \cite{conditions}.
As in \cite{wynveen} and elsewhere\cite{kauffman},\cite{others}, artificial chemistries
associated with abstracted polymers are generated consisting of strings of 
digits representing monomers. The polymers 
 undergo scission and ligation.  The parameter
$p$ controls the probability that, in a given realization, any
 possible reaction involving polymers up to a maximum length $l_{max}$ is
included in the network.  (We have regarded the small values of $l_{max}$ imposed by
computational limitations to be a flaw, but the recent discovery\cite{leslie} of ubiquitous 
`microproteins' in contemporary
organisms may suggest that our simulations of short polymers are more relevant to
prebiotic chemistry than previously thought.) Each reaction in the
network is randomly assigned one enzyme from the species present in the 
network.  The number of enzymes assigned per reaction here is different from 
the very large number of enzymes per reaction in the model of 
reference \cite{conditions} .  The choice here was made to more closely
describe the situation in real proteomes.  A more complete account of the
dependence of the model on the number of enzymes will appear later\cite{qianyi} . From the resulting chemical networks we
select, as we did previously \cite{wynveen}, those which are `viable' by which we
mean that there is at least one reaction path from a `food set' of
small polymers to at least one polymer of maximum length.
The probability that a network is viable is then found as the ratio
of the number of realizations of the network which are viable divided
by the total number of realizations.

As in \cite{wynveen} but differently
from the model described by us in \cite{intoy}, we assume here that the
system is `well mixed' and no effects of spatial diffusion are considered.
To any `polymer' ( string) of length
$L$ we attribute an energy $-(L-1)\Delta $ where $\Delta$ is a 
real number which is the bonding energy between two monomers.
The total energy $E$ of any population $\{n_{m}\}$ of polymers in which
$n_{m}$ is the number of polymers of type $m$ is $E=-\sum_{L=1}^{l_{max}}
(L-1)N_L\Delta $.  Here the $N_L=\sum_{m\ of\ length\ L} n_m $ is the same
set of macrovariables used in \cite{wynveen} and \cite{intoy}.
The total
number of polymers $N$ is $N=\sum_{L=1}^{l_{max}}N_L$.

In the simulations described here and motivated by the discussion
in the introduction, we take $\Delta$ to be negative. That means, consistent
with experiments on peptide bond formation in aqueous media, that the energy
of each bond is positive, meaning that it costs energy to make a bond.
Here $\beta$ is
assumed to be positive so that the relevant parameter $\beta\Delta <0$.
The configurational entropy associated with a coarse
grained prescription of the state given by the numbers of molecules
$N_L$ for each length $L$ between $L=1$ and $L=l_{max}$ is found by
maximizing the total configurational entropy for any set $\{N_L\}$
as given by the general Boltzmann definition $S/k_B =\ln W$ (valid whether the system
is in equilibrium or not)
with respect to the $\{N_L\}$ subject to the constraint
$\sum_{L=0}^{l_{max}} N_L = N$. Here $W$ is the number of sets of polymers
possible consistent with the set $\{N_L\}$ given that there are $b$ different
monomers available at each monomer site in all the polymers.
This is a standard problem in statistical physics \cite{landau} with the result
\begin{equation}
S/k_B=\sum_L \left [ \ln((b^L +N_L-1)!) -\ln (N_L!) -\ln ((b^L-1)!) \right ]
\label{eq:ent}
\end{equation}
for the general form of the entropy (whether the system is in equilibrium
or not. Also see our paper \cite{wynveen} for more details.)
In our simulations the polymers are not in equilibrium but, in addition to the nonequilibrium distributions calculated from kinetics,  we also calculate the
distributions $\{\overline{N_L}\}$ associated with local equilibrium and
equilibrium with a temperature bath at temperature $T$ continuously
during the simulations.  Those distributions are found by maximizing
the entropy given above with respect to the variables $\{N_L\}$ subject to
different constraints depending on whether the equilibrium attained is
that resulting from an open system in contact with a thermal bath at
fixed temperature $T=1/k_B\beta$ or, on the other hand, is
sufficiently isolated to permit it to attain local equilibrium  consistent
with the current value of the total energy $E$.  In the former, open,
case, the maximization is carried out taking account of the constraint
with a Lagrange multiplier $\mu$ which is the chemical potential
whereas in the second, closed or local case, the maximization
is carried out taking both constraints (energy and particle number)
into account with Lagrange multipliers
$\mu$ and $\beta = 1/k_BT$, which are determined from the inputs $N$ and $E$
(as described in detail in
reference \cite{conditions}).
In both cases, $\overline{N_L}$ is found
to have the form
\begin{equation} 
\overline{N_L}={{b^L-1}\over{\exp(-\beta\mu-\beta\Delta(L-1))-1}}
\label{eq:avNL}
\end{equation}
This expression is formally equivalent to that found for a bose gas in
elementary quantum statistical mechanics although this model has no
explicitly quantum features and no quantum effects are suggested or
implied . See references \cite{wynveen} and \cite{landau}. 
In the 
`Gibbs limit' in which $b^L >> N_L >> 1$ all these expressions reduce
to the familiar classical Boltzmann equilibrium formulae as discussed
in \cite{landau} and the appendix to this paper, for example. 
That limit applies in the present 
case for proteins in most biological contexts.   But for RNA, for which $b=4$, it is only valid for
large $L$ and corrections to the Boltzmann equilibrium formulae are
nonnegligible for many of our simulations.
To determine the isolated equilibrium state we compute $\beta$ and
$\mu$ from the known energy $E$ and polymer number $N$ by solution
(on the fly during the simulations) of the equations
\begin{equation}
E=-\sum_{L=1}^{l_{max}}
(L-1) \overline{N_L}\Delta
\label{eq:Econd}
\end{equation}
and
\begin{equation}
N=\sum_{L=1}^{l_{max}}\overline{N_L}
\label{eq:Ncond}
\end{equation}
whereas to determine the equilibrium state resulting from 
equilibrium with a temperature bath we fix $\beta$ and determine
$\mu$ by solution, again on  the fly, using the equation \ref{eq:Ncond}
. Note that, in this formulation and throughout the paper, references
to `equilibrium' refer to  maximization of the configurational
entropy, constrained only by the number of polymers and the external 
temperature (when it is specified) or the total energy.  Thus we 
mean maximization given that all the polymer configurations in the 
model are `accessible', whereas in some work in chemical statistical
mechanics, further constraints are applied, arising from the assumption
that some states are kinetically `inaccessible'.  Such kinetic 
inaccessibility does occur in our models, through the kinetic model
described next, but it does  not enter our definition of the equilibrium 
states.   

During the dynamics simulation, the temperature  enters the 
dynamics through the factors $k_d$  in the  following kinetic Master equation 
 (equation (\ref{eq:master})).
\begin{eqnarray}
\label{eq:master}
  dn_l/dt = & \sum_{l',m,e} [v_{l,l',m,e} (-k_dn_ln_{l'}n_e+k_d^{-1}n_mn_e) \nonumber \\*
              & +v_{m,l',l,e}(+k_dn_mn_{l'}n_e-k_d^{-1}n_ln_e)]. 
	      \end{eqnarray}
Here $n_l$ is the number of polymers of species $l$,
$v_{l,l',m,e}$ is proportional to the rate of the reaction
$l+l'{{e}\atop{\rightarrow}} m$,$e$ denotes the catalyst, $l$ and $l'$ denote
the polymer species combined during ligation or produced during
cleavage, and $m$ denotes the product of ligation or the reactant
during cleavage. 
This model
for the dynamics, defined by the Master equation,
is stochastically simulated using the Gillespie algorithm\cite{gillespie}. 
A parameter $p$ (in $[0,1]$) controls the sparsity of 
reactions in the network. With probability $p$, each reaction rate
has a finite value  $v_{l,l',m,e}$ or $v_{m,l',l,e}$ but the rate is 
zero with probability $(1-p)$ for each possible reaction. The values of 
$v_{l,l',m,e}$ or $v_{m,l',l,e}$ are fixed (from a uniform distribution in 
$[0,1]$) during the dynamical simulations
but the values of the parameters $k_d$ are not.  The latter are fixed by the
detailed balance condition
\begin{equation}
k_d^2 = \overline{n_m}/(\overline{n_l}\ \overline{n_{l'}})
\label{eq:kdeq}
\end{equation}
where, in the simulations reported here, the equilibrium distributions 
$\{n_l\}$ in the last expression are always taken to be those associated 
with equilibrium with an external thermal bath with a fixed parameter $\beta$.
We started all the simulations reported in this paper with a 
`food set' of 500 randomly selected (from the $b(b+1)$ available) monomers and
dimers. Some results do depend on the choice of the starting food set.
However we are assuming, as in much work on the origin of life, that 
the problem
is to understand how lifelike systems emerge from collections of small,
interacting molecules, so that starting from monomers and dimers will 
give relevant insights.   
During the simulations, the simulated systems
are `fed' by maintaining the population of dimers and monomers above a specified
minimum usually taken to be 500 . (Thus the system is `open'\cite{blokhuis}.)
The system is continually driven toward equilibrium with the external
thermal bath but many simulated systems do not achieve  either
local equilibrium or equilibrium with the external bath because of the kinetic
blocking imposed by  $p< 1$ and because of the `feeding'.  As in our previous
work, including that described in \cite{conditions} and \cite{wynveen}, 
we assume that lifelike chemical systems
will be metastable  states far from equilibrium and select and count such
states to obtain a quantitative indication of how likely our models are to
result in lifelike states. 

As in \cite{intoyhalleypopulations} and \cite{conditions} we compute
 two Euclidean distances $R_L$ and $R_T$ in the $l_{max}$ dimensional
space of sets $\{N_L\}$ which characterize how far the system of 
interest is from the two kinds of equilibria described above:
\begin{equation}
R_L=\sqrt{\sum_{L} (N_{L}- \overline{N_L(\beta(E,N),\mu(E,N))})^2}/(\sqrt{2}N)
\label{eq:RL}
\end{equation}
for distance from the locally equilibrated state and
\begin{equation}
R_T=\sqrt{\sum_{L} (N_{L}- \overline{N_L(\beta,\mu(\beta,N)}))^2}/(\sqrt{2}N)
\label{eq:RG1}
\end{equation}
for distance from the thermally equilibrated state. 
The normalization factors in these equations differ from that in 
equation \ref{eq:RG} because, in equation \ref{eq:RG} we took account
of polymer dilution and used experimental data on polymer volume density
instead of total polymer number $N$. The normalizations are chosen in 
each case so that the resulting quantities $R_T$ and $R_L$ lie in the
interval $[0,1]$ permitting the model values to be compared with the
experimental ones. Alternative measures of the degree of disequilibrium in the
context of study of polypeptide systems have been proposed\cite{stopnitzky} 
and
we have used alternative formulations in references \cite{wynveen} and 
\cite{intoy} .  The formulation used in this paper and in \cite{intoyhalleypopulations} and
\cite{conditions} has the advantage of discriminating
between local equilibrium, which would be achieved by the system in isolation
and the global or thermal equilibrium with an external thermal environment,
which would be eventually achieved if the system were in contact with
an external, equilibrated `bath'. The latter distinction has provided
valuable insights into the nature of the nonequilibrium states found 
in our quench simulations. A similar Euclidean measure of disequilibrium
in the context of prebiotic evolution was also suggested in reference \cite{baum}.
More details of the simulation methods are
described in \cite{conditions}. 

The physical significance of $R_L$ and $R_T$ is that $R_L$ measures
the distance in the population space of the current simulation point
from the point where it would be if the system had self equilibrated 
consistent with its total internal energy $E$ but not with any external
temperature bath, whereas $R_T$ measures the distance in the population space of the current simulation point
from the point where it would be if the system had equilibrated to an
external temperature bath at temperature $T=1/k_B\beta$.  One generally
expects self equilibration to occur faster than equilibration to an
external temperature bath. The temperatures of the corresponding equilibria
are often different in condensed matter systems, for example electron
temperatures, both in plasmas \cite{gonzalez} and in solids (eg \cite{perrez}) can be different
from ion or lattice temperatures respectively and the internal temperature
associated with the distribution of nuclear spin directions in NMR 
experiments\cite{abragam} can be different from the lattice temperature (and negative
in the latter case.)
  
In the results cited in section   \ref{sec:Results}  the code implementing this model
 was modified to permit an abrupt change in the parameter $\beta\Delta$ 
 during the simulation of systems in contact with an external thermal bath.
 In the report of results which follows, we change the value of $\beta\Delta$
 from a small negative value to a large one.  The values are negative
 because the free energy of bond formation of peptide bonds in water is
 negative \cite{martin1998free} as noted above and the choice of small to large negative
 values will correspond, in the case that $\Delta$ does not change, to 
 a quench from high to low temperature. We thus refer in the discussion
 to quenches from high to low temperature, but note that the relevant parameter
 in the model is the product $\beta\Delta$ (the two factors always occur 
 together) and a similar change in that parameter might be induced by
 altering $\Delta$ for example by a rapid change in pH \cite{pHdependence}

\section{Hypotheses}\label{sec:Hyp}

If the model described in the preceding section adequately describes the
coarse grained features of the relevant prebiotic chemistry, then the
conjectures concerning the origin of prebiotic chemistry in quenches
of interacting amino acids from high to low temperature described in
the introduction imply that we should expect  the
following simulation results within the model:

1. Running at initial high temperature $\beta\Delta_i$ and then quenching
to $\beta\Delta_f$, one should find a minimum, during the low
temperature part of the run,in
\begin{equation}
R_T(\beta\Delta) =\sqrt{\sum_L (N_L(\beta\Delta_f)-\overline
N_L(\beta\Delta))^2}/\sqrt{2N}
\label{eq:RThalfcold}
\end{equation}

as a function of $\beta\Delta$ at $\beta\Delta=\beta\Delta_i$.
Here $N_L(\beta\Delta_f)$ is
the value of $N_L$ found from the kinetic simulation with external
temperature $\beta\Delta_f$ and $\overline N_L(\beta\Delta))$
is the equilibrium expression for $N_L(\beta\Delta )$ at the temperature
$\beta\Delta$. 
We do consistently find such a minimum, though the minimum value
of $R_T$ varies as discussed later. We denote the value of $\beta\Delta$
at the minimum by $\beta\Delta_{min}$ .  Then the hypothesis states
that $\beta\Delta_{min}=\beta\Delta_i$.  We will present numerical 
evidence that, within the model, this is approximately true when
$p$ is sufficiently large.

The significance of this is that if
an experimental system, such as one of the proteomes we
studied previously, exhibits such a minimum in $R_T(\beta\Delta)$
with $N_L(\beta\Delta_f)$ replaced by the experimental values
of $N_L$ then the $\beta\Delta$ at which a minimum in $R_T(\beta\Delta)$
occurs is a signature of the temperature at which the system
formed before quench. Thus, since the proteomes had such a
minimum at around 400K, our simulations would support the
idea that the system formed from a quench at a high
temperature around 400K. (See Fig. \ref{fig:RTfuncdb}.)

2. The high temperature at which equilibrium should occur most
easily should be the one which gives
\begin{equation}
(1/l_{max})\sum_L  \delta\overline{N_L}/\delta L(\beta\Delta ) =0.
\label{eq:dNLdL0}
\end{equation}
(Here $\delta \overline{N_L}$ and $\delta L$ denote finite increments.
In implementation we take $\delta L =1$.) 
We call the solution to this equation $\beta\Delta_{flat}$.

To explain the motivation for this hypothesis further, we refer
to the discussion of Fig. \ref{fig:MartStdDev} in the introduction 
to this paper.  There we pointed out that very small changes in
the assumed binding energy of peptide bonds produced very large 
changes in the equilibrium polymer length distribution if the temperature
were near ambient. So at those, low, temperatures the equilibrium 
distribution would be very likely to be either almost all monomers or
almost all polymers of maximum length (the blue and orange bars in 
the Fig.). If the lengths were distributed in that fashion in 
prebiotic conditions then the kinds of interactions possible would
be highly constrained if the system were near equilibrium.  That,
in turn would greatly slow the rate at which polymer configurations
were naturally explored by the kinetics, slowing, in turn, the natural
search (in a dynamic network as envisioned in \cite{taran}) 
for the rare combinations which turn out to be lifelike.
However the actual distribution of protein lengths in prokaryotes
is usually like that shown for one of them by the purple bars in 
Fig. \ref{fig:MartStdDev} and such a distribution
can be envisioned to permit the kind of natural search
required. Therefore we sought, in formulating hypothesis 2, a
condition on the initial temperature which would require that the
equilibrium length distribution be reasonably flat so that, if the system
were near equilibrium in the high temperature phase before quench,
it would be capable of the kind of natural search envisioned in 
\cite{taran} and by us. Equation (\ref{eq:dNLdL0}) realises that aim
by requiring that the average discrete derivative of $\overline{N_L}$ with
respect to $L$ be zero.  However if the  temperature is 
high and gives a relatively flat distribution as required by 
equation (\ref{eq:dNLdL0}), then, in the high temperature phase of the 
simulation before quench the system will most easily come close
to equilibrium. 
That would say that
if, at high temperatures, one computes $R_T(\beta\Delta_i)$ in our model 
(with BOTH $\beta\Delta$ arguments in the definition =$\beta\Delta_i$)
then one should find the smallest $R_T$ at the value of
$\beta\Delta_i$ for which \ref{eq:dNLdL0} is true, as postulated in
Hypothesis 2.

3.  The most lifelike states after quench should occur when the initial high
 temperature state coincides with the minimum described in 2. 
 To characterize this prediction quantitatively requires a more detailed
 statement of the definition of `lifelike states' as we will discuss in the
 next section. 

4.  The solution to $(1/l_{max})\sum_L \delta \overline{N_L}/\delta L(\beta\Delta ) =0$
will be at $\ln b=-\beta\Delta$ in the large b limit and that is the 
appropriate limit for the proteome systems.

\section{Results} \label{sec:Results}

\begin{figure}
\includegraphics[width=3.5in]{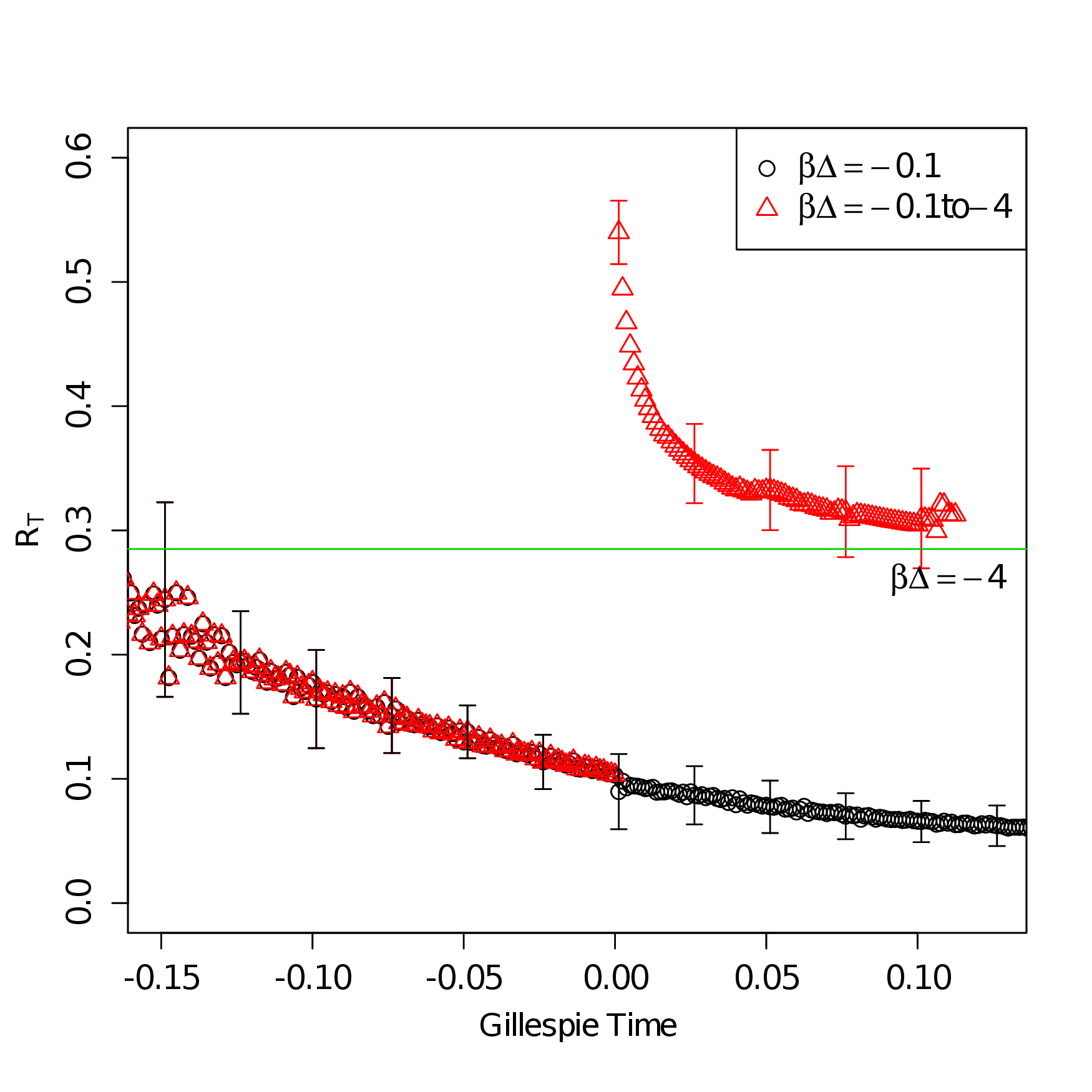}
\caption{\label{fig:dynamicsexample1}Average over 30 runs on different networks of the value of $R_T$ versus Gillespie simulation time. Gillespie times have been shifted by a constant
so that the quench always occurs at Gillespie time =0.  Some data from times earlier
than Gillespie time (as shifted) less than -0.15 which arise from noisy initial transients 
have been omitted. Error bars indicate standard deviations over the ensemble. Total simulation 
time 
was fixed to be $3 \times 10^6$ reaction steps and quench was applied at $3 \times 10^5$ reaction step.  Parameters are $p=0.1280$, $b=4$, $l_{max}$=7.
}
\end{figure}

Before describing results of tests of the hypotheses of section \ref{sec:Hyp}
we show an example of how $R_T$ characteristically behaves in Gillespie simulation 
time during a subset of our simulated quenches. In Fig. 
\ref{fig:dynamicsexample1}) the black circles show average values of 
$R_T$ over 30 runs on the same network in a simulation at a high temperature
($\beta\Delta = -.1$),
 well above ambient temperature. The high temperature systems start far 
from the equilibrium associated with the external temperature 
,because all simulations start with just a randomly selected food set,
but quite
rapidly approach a state near equilibrium. Here, as in most cases 
and even without
significant kinetic blocking, our simulations do not go all the way
to equilibrium because they are continually being `fed' by maintenance
of the food set population. 
The red triangles show what happens when the simulations
are repeated starting with the same high temperature  with a quench at time zero to a lower temperature
($\beta\Delta = -4.$).   The systems attain a high value of $R_T$ after
quench because the equilibrium point has moved in the species space,
but the instantaneous populations have not.  The systems then move closer
to equilibrium at the lower temperature but do not achieve it.  To
see if the quench has produced systems farther from equilibrium than a 
simulation entirely at the lower temperature would do, we also show 
,in the green symbols, the results
of a simulation on the same systems when the external temperature remains
at the low value ($\beta\Delta = -4.$) throughout . The green curve 
flattens at a higher value of $R_T$ than the high temperature one
(black symbols) but below the quenched values (red symbols) indicating
that the quench has produced systems farther from low temperature 
equilibrium than would simulating from the lower temperature from
the start. However the effect is relatively small  in these examples.
A more comprehensive set of data evaluating the effect of quenching
on disequilibrium appears in Fig. \ref{fig:changeofRT}.

\begin{figure}
\includegraphics[width=3.5in]{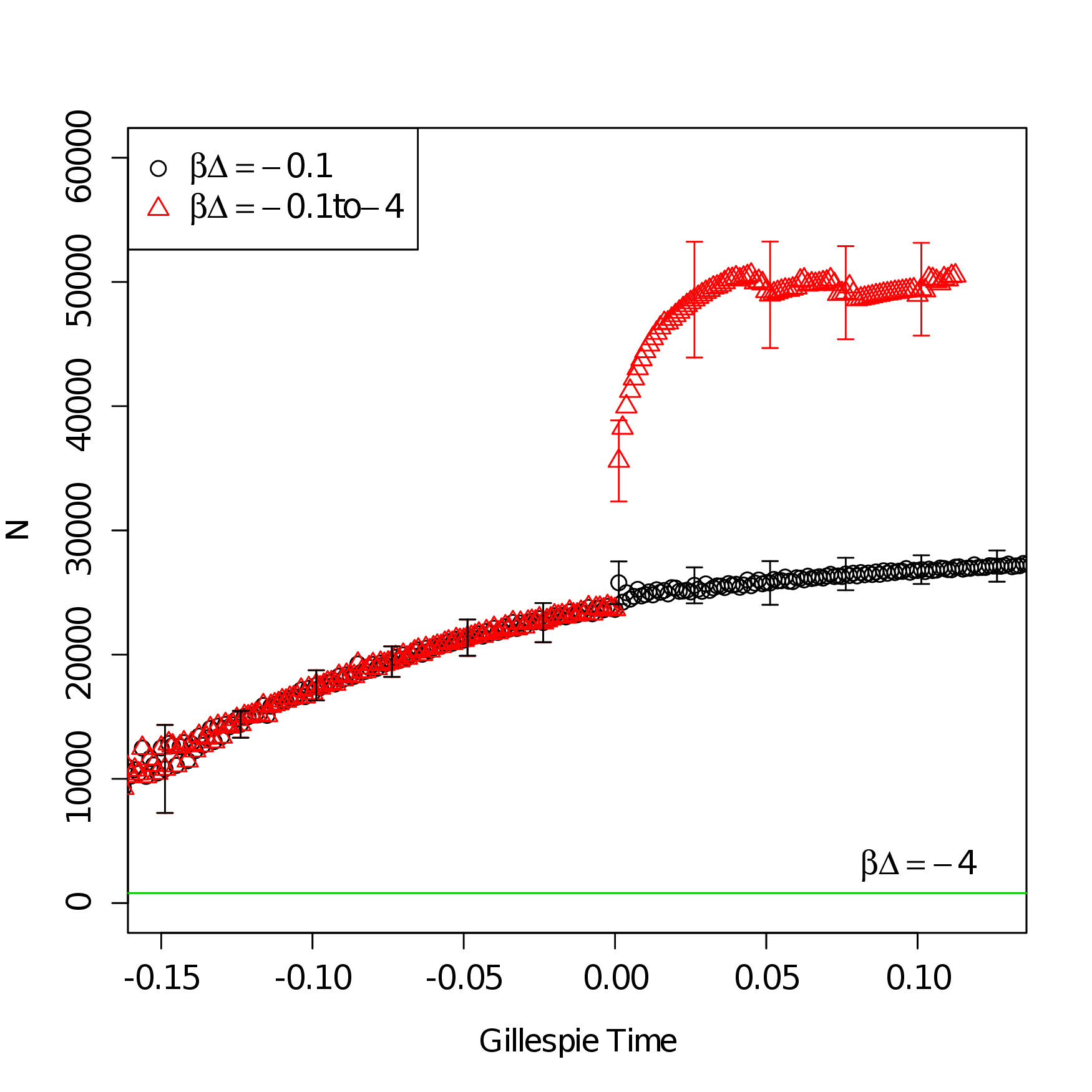}
\caption{\label{fig:dynamicsexample2}
Average over 30 runs on different networks of the value of N, the total number of polymers, versus Gillespie simulation time. Error bars indicate standard deviations over the ensemble. Parameters are $p=0.1280$, $b=4$, $l_{max}$=7.
}
\end{figure}

In Fig. \ref{fig:dynamicsexample2} we show the 
total polymer populations as a function of the number of reaction 
steps for the
same ensemble of realizations of the model.  
The quench has a much bigger effect on this variable:  The final 
number of polymers is much larger throughout times after the quench
than it is at low temperature after a quench and also bigger
than the final number from the run at the  higher temperature throughout.
Of equal interest, the quenched systems continue to grow rapidly 
after quench whereas both the low and high temperature runs show very
little population growth near the end of the runs. 
Part of the growth in the total number of polymers after quench is 
due to the increased rate of scission relative to ligation at the
lower bath temperature. 

In these examples we thus have preliminary evidence that three properties
deemed lifelike, namely disequilibrium (measured by $R_T$), population 
size and population growth rate, are enhanced by quenching from
high temperatures. Note that, in the model, the temperature always
occurs in the combination $\beta\Delta$ so that identical results
could also be achieved by a sudden change in $\Delta$ as, for example,
might be achieved by a sudden change in pH.   
 
The `error bars' in the figures indicate standard deviations in the distribution
of results over this ensemble and suggest, consistent with our detailed 
results, that some realizations can experience much higher
values of disequilibrium and growth rate after quench.

\begin{figure}
\includegraphics[width=3.5in]{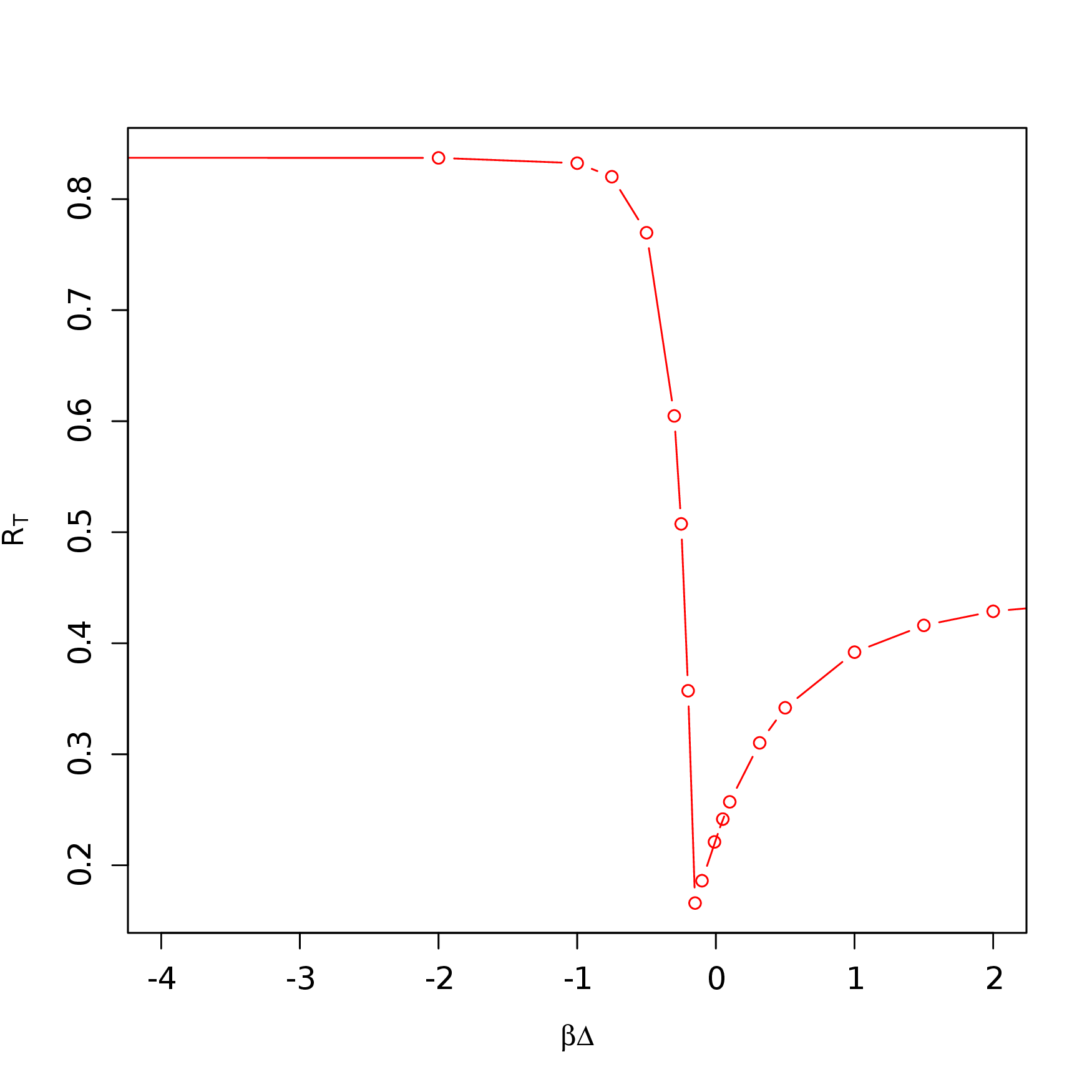}
\caption{\label{fig:RTvsdbmodelb} $R_T$ from equation 10 calculated from a dynamic simulation with $b = 4$, $l_{max} = 7$, $p = 0.1280$, $\beta\Delta_i = - 0.1$ and $\beta\Delta_f = - 4$. Compare \ref{fig:RTfuncdb} which was obtained from proteome data. 
} \end{figure}

In the following data we show results bearing on the validity of the
hypotheses we consistently fixed the parameters $\beta\Delta_f=-4.,l_{max}=7$
and , where not specified otherwise, $b=4$.  
The value of $\beta\Delta_f$ was chosen
to approximately match the value for peptide bonds under ambient 
conditions. The values $b=4$ and $l_{max}$ are approaching the limit 
imposed by 
constraints on available simulation resources.  Networks had
reactions with 1 enzyme per reaction.  Results are presented for 
a series of $p$ values in the model, since the $p$ value turns out 
to be significant. 

\begin{figure}
\includegraphics[width=3.5in]{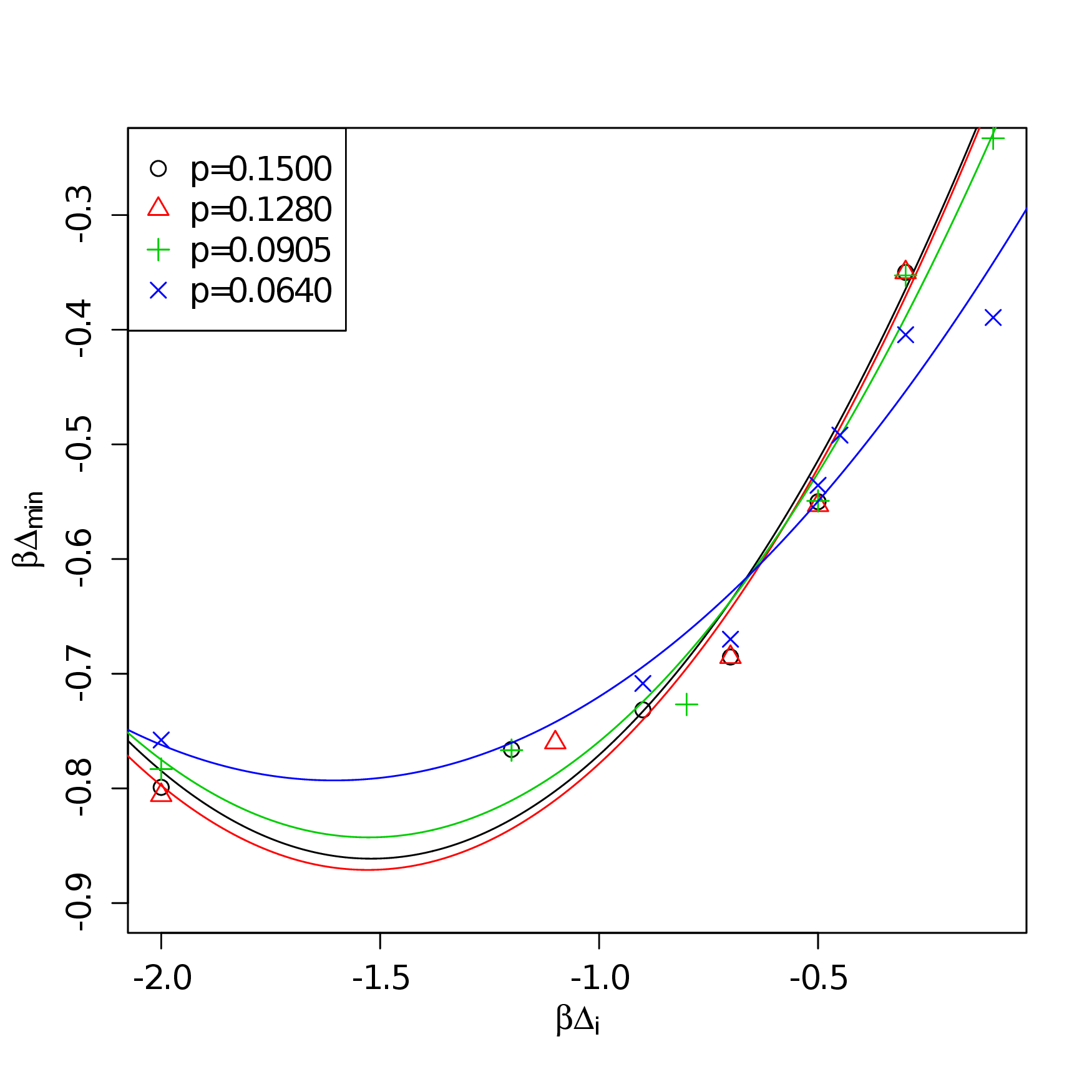} 
\caption{\label{fig:dbminvsdbipolynomialfit} 
Simulated values of $\beta\Delta_{min}$ versus $\beta\Delta_{i}$ for various $p$ values. 
The final temperature are set so that  $\beta\Delta_{f}=-4$. Each data point comes from 
an average over 40~500 realizations. The simulation used values $b=4$, $l_{max}=7$. 
The smooth curves show the result of fitting these data, for each value of $p$
to a quadratic function of $\beta\Delta_{i}$ as explained in the text.}\end{figure}

Regarding hypothesis 1. We show values of $\beta\Delta_{min}$ as a function
of $\beta\Delta_i$ for a series of $p$ values in Fig. \ref{fig:dbminvsdbipolynomialfit}.
The data get closer to the 
hypothesis as $p$ increases and $\beta\Delta_i$ becomes less negative 
corresponding to higher temperatures and a more connected chemical
network.

To explore hypothesis 1 further we 
fit the function
$\Delta\beta_{min}=f(\Delta\beta_{i})= f(0) +f'(0)\Delta\beta_i
+(1/2) f''(0) (\Delta\beta_i)^2 $, ie the first terms in a Taylor series in
$\Delta\beta_i $ to the data in Fig. \ref{fig:dbminvsdbipolynomialfit}
for five $p$-values between .05 and .3 with results indicated by the smooth curves in
the Fig.. Hypothesis 1 states that we will find $f(0)=0,f'(0)=1, f''(0)=0$
. Results are shown in  Table \ref{tab:fitparams}.  The hypothesis is
quite nearly confirmed for the larger values of $p$.

\begin{table*}
\caption{Polynomial fit parameters for the data in Fig. \ref{fig:dbminvsdbipolynomialfit}}
\label{tab:fitparams}
\begin{tabular}{|c|c|c|c|}
\hline
p & $f(0)$ & $f'(0)$& $f''(0)/2$ \\ \hline \hline
0.1500&        -0.09019 +- 0.04496&	1.01408 +- 0.10566&     0.33341 +- 0.04801 \\ \hline
0.1280&	      -0.09674 +- 0.04552&	1.01314 +- 0.11502&     0.33139 +- 0.05181 \\ \hline
0.0905&	      -0.13961 +- 0.04393 &      0.92138 +- 0.10938 &    0.30185 +- 0.04991 \\ \hline
0.0640 &      -0.27957 +- 0.04638 &      0.63970 +- 0.11963 &    0.19924 +- 0.05271  \\ \hline
0.0226 &      -0.63764 +- 0.01819 &      0.12222 +- 0.04276 &     0.03128 +- 0.01943  \\ \hline
\end{tabular}
\end{table*}

\begin{figure}
\includegraphics[width=3.5in]{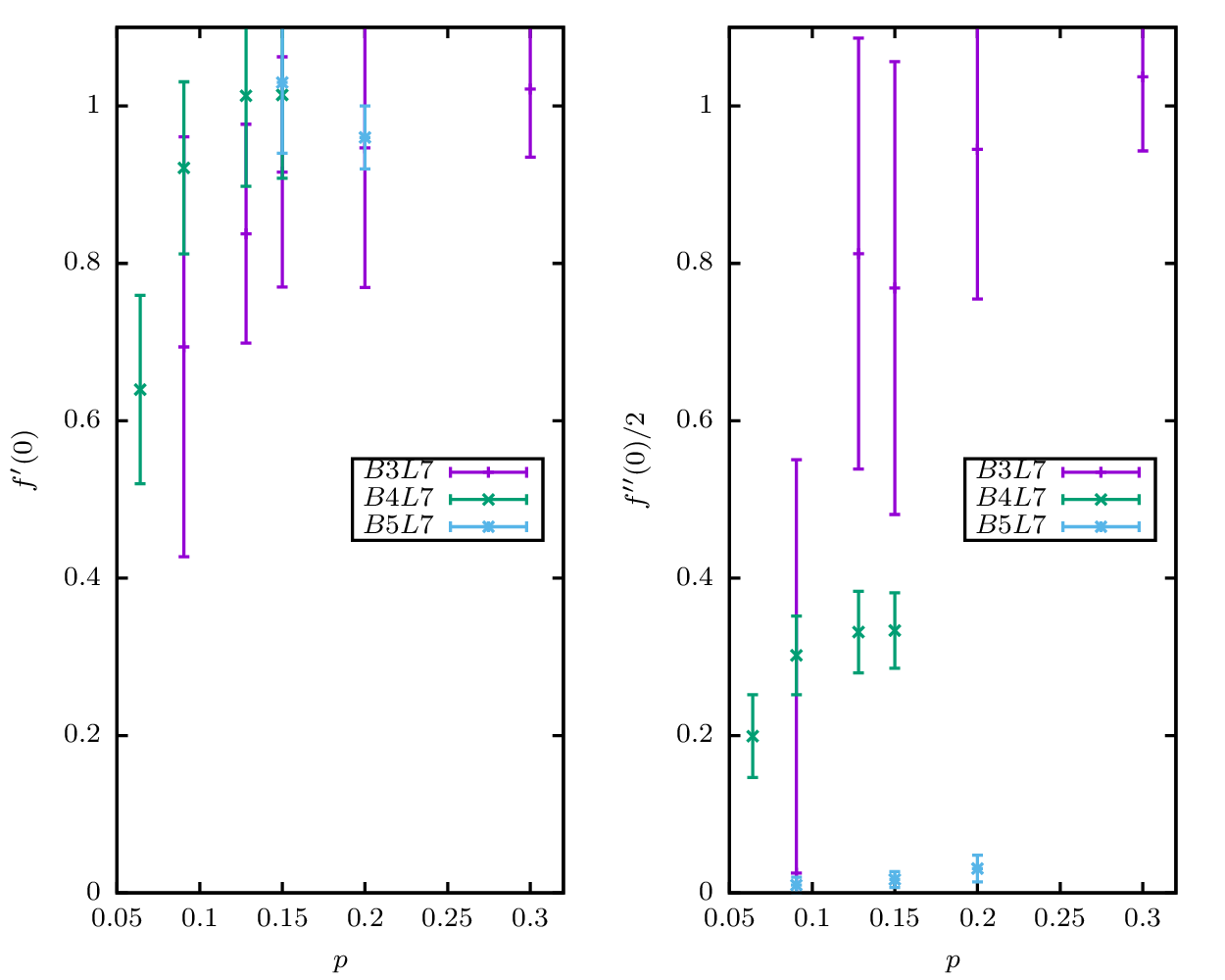}
\caption{\label{fig:pvsslope2}
Values of the slope (left) and curvature (right) of  
fits to the data in Fig. \ref{fig:dbminvsdbipolynomialfit} and similar data
for $b=2,4$ and $5$ to  quadratic functions of 
$\beta\Delta_{i}$ as a function of $p$ as explained in the text. Error bars indicate standard errors.
}
\end{figure}

To explore the dependence of this result on $b$ we carried out a similar 
analysis 
for $b=2,3,$
and $5$ and show the results for the fitting parameters as a function of $p$ 
for five $p$-values between .05 and .30 
with results shown in Fig. \ref{fig:pvsslope2} . As $b$ gets larger
the quadratic term gets smaller, indicating that the hypothesis works 
for larger and larger values $\beta\Delta_i$ and consistent with the 
suggestion that   at the biologically relevant value of $b=20$ it will 
continue to be valid up to values of $\beta\Delta_i =-\ln 20 $.
  
Note that when $b$ gets larger, even a
relatively small $p$ gives a slope near 1.  We conclude that for large 
$b$ systems such as the one in the proteomes with $ b=20 $ it’s very
likely that $\beta\Delta_i = \beta\Delta_{min}$ in agreement with 
hypothesis 1.  Thus our previous inference that the value for 
$\beta\Delta_{min}$ taken from the proteome data was an indicator of
the temperature from which the proteome had been quenched on the 
early earth is consistent with our model.

\begin{figure}
\includegraphics[width=3.5in]{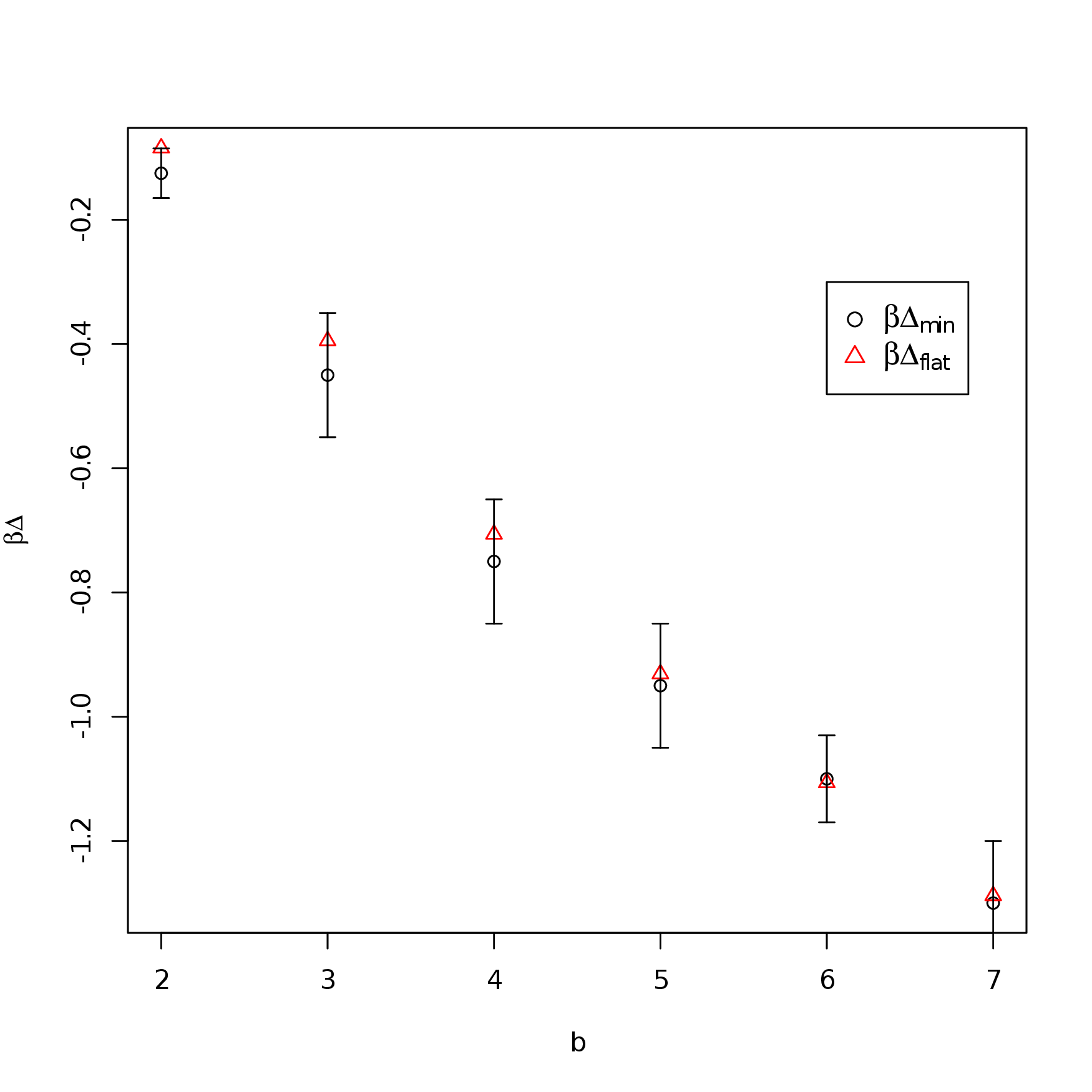}
\caption{\label{fig:N-SCompare} 
$\beta\Delta_{flat}$ and $\beta\Delta_{min}$ versus $b$. $\beta\Delta_{min}$ is determined by running simulations with a series of temperature $\beta\Delta_{i}$ and finding minimum $R_{T}(\beta\Delta_{i})$. The error bars are determined when a $\beta\Delta_{i}$ gives a $R_{T}(\beta\Delta_{i})$ that is within $10\%$ difference of $R_{T}(\beta\Delta_{min})$. 
.}
\end{figure}

\begin{figure}
\includegraphics[width=3.5in]{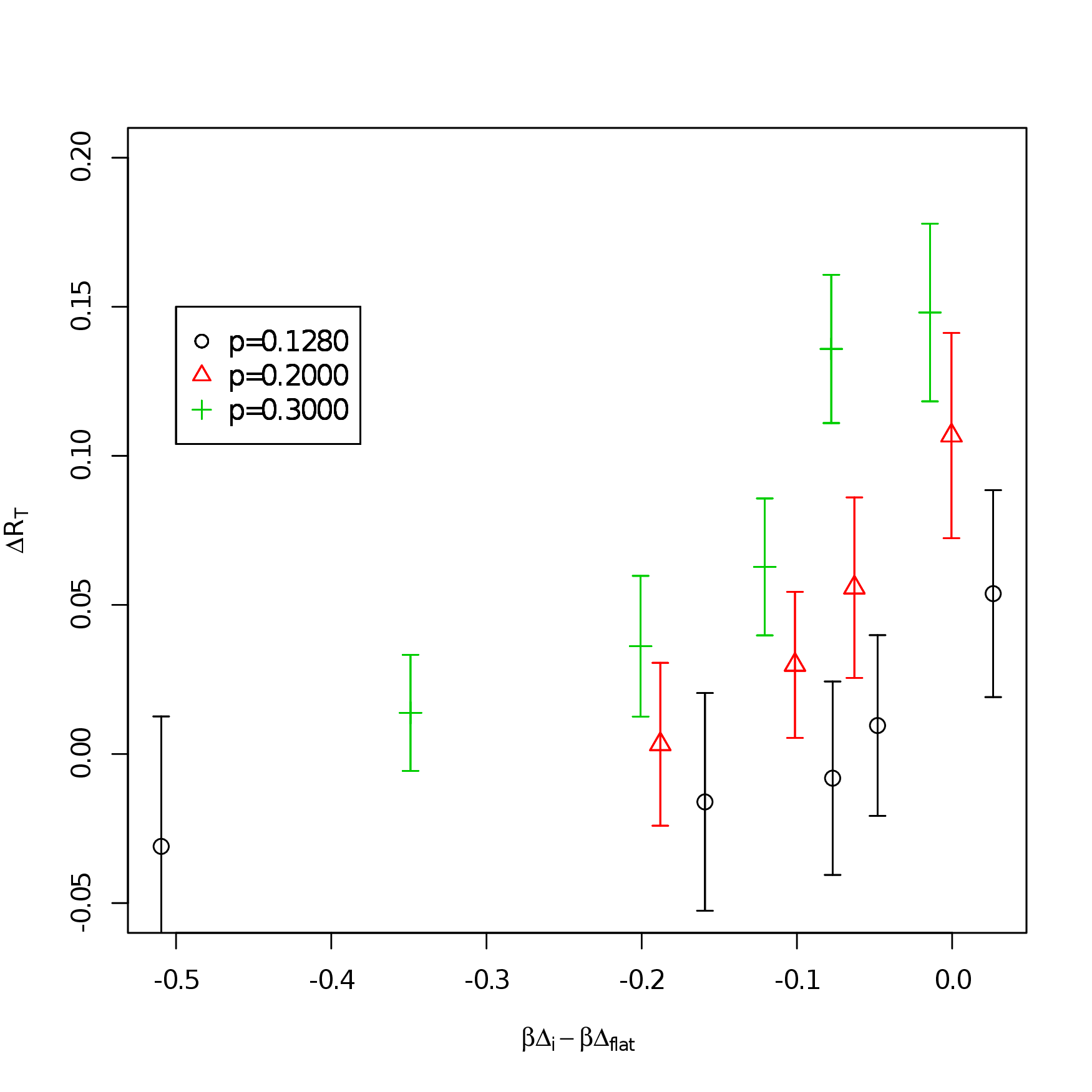}
\caption{\label{fig:changeofRT} Enhancement of $R_T$ as a function of $\beta\Delta_{i} - \beta\Delta_{flat}$ for various values of $p$. $\beta\Delta_f=-4$,$l_{max} = 7$, $b = 4$. Each point is from between 20 and 50 simulations with smaller numbers
of simulations for larger values of $p$.  Error bars indicate standard deviations. 
}
\end{figure}

To test hypothesis 2 we calculated solutions to equation 
\ref{eq:dNLdL0} numerically for values of b=2,3,4,5,6,7.  Hypothesis
2. states that the value of $\beta\Delta_i$ at which 
\begin{equation}
R_T(\beta\Delta_i) =\sqrt{\sum_L (N_L(\beta\Delta_i)-\overline
N_L(\beta\Delta_i))^2}/\sqrt{2N}
\label{eq:hotRT}
\end{equation}
is minimum (termed 
$\beta\Delta_{min}$) should be at the same value of $\beta\Delta_i$, termed $\beta\Delta_{flat}$ 
at which 
(\ref{eq:dNLdL0}) is true. (Note that, in
equation \ref{eq:hotRT}, BOTH arguments $\beta\Delta$ are at 
$\beta\Delta_i$ (the 'hot' value) unlike equation \ref{eq:RThalfcold}.
Hypothesis 2 is a statement only about the `hot' phase.) There is a complication here because $\overline
N_L(\beta\Delta_i)$ depends on $\mu\beta$ which, in turn,  depends on the
total number $N$ of polymers present in the current state of the system.
To determine $\beta\Delta_{min}$ we used that current value of $N$ at each 
value of $\beta\Delta_i$ in equation (\ref{eq:hotRT}) and then used the
$N$   value at $\beta\Delta_{min}$ to give a value for $\beta\mu$ to use
in solution of equation (\ref{eq:dNLdL0}) for $\beta\Delta_{flat}$.  
The values of $\beta\Delta_{min}$ and $\Delta\beta_{flat}$ are compared for  $b$=2,3,4,5,6,7 and 
$p=0.0226,l_{max} =7$ in Fig. \ref{fig:N-SCompare}. The  trends in the two 
quantities are the same and the values are close to one another
but not identical.  As $b$ gets larger they get closer.  We conclude
that hypothesis 2 is likely to be a very good approximation when 
$b$ is large, as in the proteomes.

Regarding hypothesis 3, we first tested for the enhancement of $R_T$,
termed $\Delta R_T$ in 
the cold phase, relative to the value obtained in the same network with
the same starting conditions when the temperature was low throughout the 
run. The hypothesis states that $\Delta R_T$ should be largest when 
$\beta\Delta_i-\beta\Delta_{flat}$ is zero.  We show results for
three $p$ values in Fig. \ref{fig:changeofRT}. In all cases, the
enhancement rises sharply as $\beta\Delta_i\rightarrow \beta\Delta_{flat}$
confirming hypothesis 3 for the lifelike property of disequilbrium.
Qualitatively, the enhancement occurs because at the high
temperature, the systems get close to the internal equilibrium population
distribution 
imposed by their total energy and retain a distribution of polymer
lengths close to that value after quench. The enhancement increases with
$p$, possibly because the high temperature phase equilibrates more effectively
before quench as $p$ increases. However for very large $p$ the effect of 
the stabilization of nonequilbrium states by quenching is expected to
become less effective because the denser network will allow equilibration
even in the low temperature state. 

\begin{figure}[htbp]
 \centering
 \includegraphics[scale=0.5]{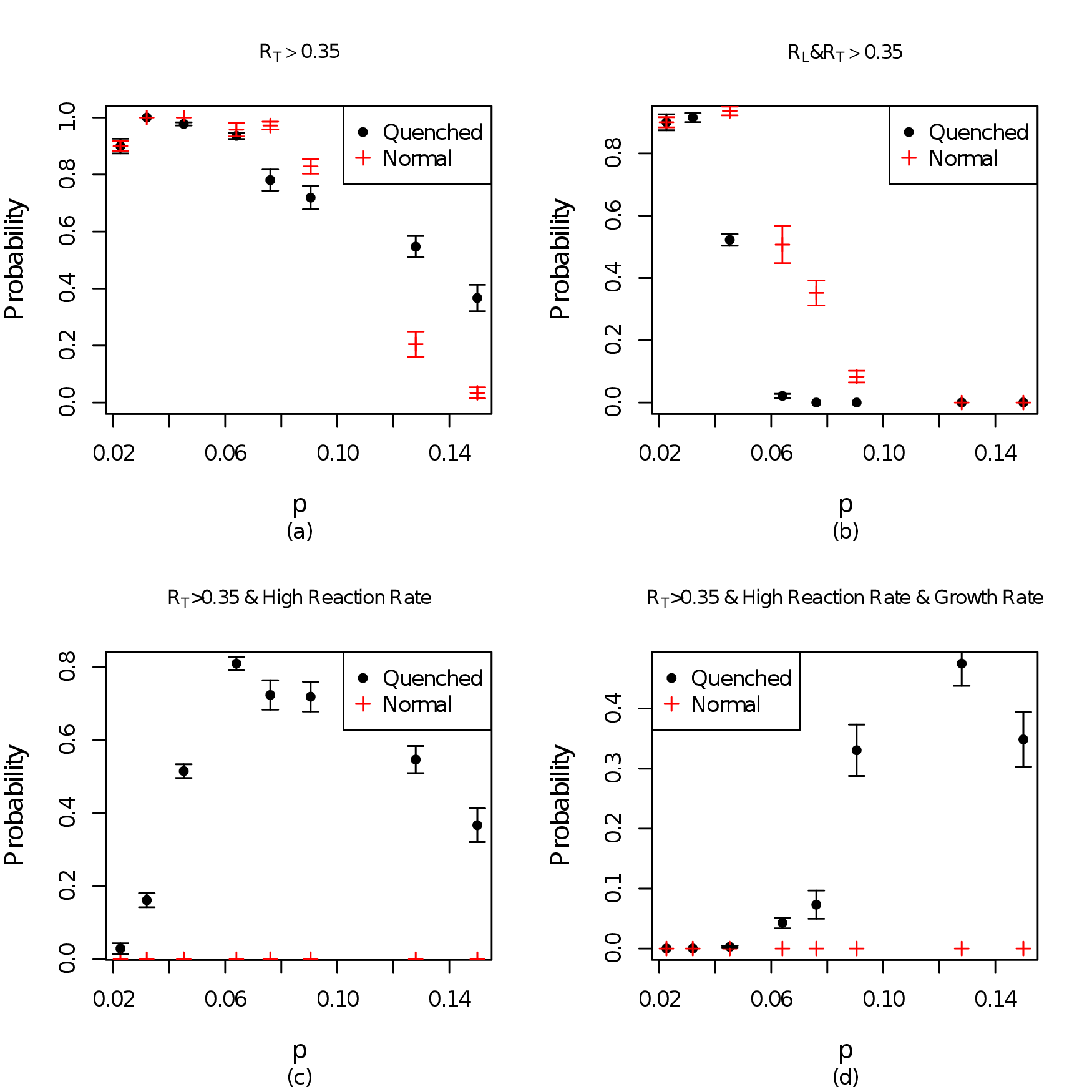}
 \caption{\label{fig:lifelikefilters}(a) Probability of finding   $R_T > 0.35$
 with (Quenched) and without (Normal) quenching, (b) same with 
 $R_L$ and $R_T > 0.35$, (c) Probability of finding $R_T > 0.35$  and a  
 high reaction rate , (d) Probability of $R_T > 0.35$, a   high reaction rate and a high growth rate. The threshold for the reaction rate cut off is 10 reactions per unit of Gillespie time per polymer. The growth rate cut off eliminates those systems for which $d\ln N/dt <1$ in units of
 inverse Gillespie time. For the quenched simulations, $\beta\Delta_i = 
 \beta\Delta_{flat}$ and $\beta\Delta_f = -4$. For the unquenched 
 simulations $\beta\Delta_i = \beta\Delta_f = -4$. $b=4$, $l_{max} = 7$, for 
 both cases. From data  on 50 to 700 realizations with more simulations
 for the smaller values of $p$. Error bars show standard deviation from the 
 average over the simulations for each parameter set.}
 \end{figure}

To study the effects of quenching on other, possibly lifelike
, properties we applied a series of filters to the ensembles of systems
obtained by quenching and show results in Fig. \ref{fig:lifelikefilters}.
(Quenching from the high to low environmental temperature does not enhance
$R_T$ significantly for the range of $p$ values used here,
but does enhance $R_T$ at larger $p$ values as indicated in
Fig. \ref{fig:changeofRT}.)

 In part c. of Fig. \ref{fig:lifelikefilters} we 
 show results of imposing an additional filter which excludes results in
 which the reaction rate per polymer in the final steady state is below a fixed value. 
 This `dynamics filter' is different than the one imposed on 
 our results in references \cite{conditions,wynveen,intoy}. 
To make the cut we require that the total number
 of reactions per polymer in the final steady state divided by the gillespie time 
 elapsed during that steady state part of the run be larger than
a fixed value  which, in the data displayed in 
Fig. \ref{fig:lifelikefilters}, we chose to be 10 reactions per unit
of Gillespie time. (Roughly, one unit of Gillespie time corresponds to
the average rate of ligation and scission.) There is a very
 large enhancement of reaction rate due to quenching.

Finally, in
 part d of Fig. \ref{fig:lifelikefilters} we show the effects of further
 filtering to isolate the final steady states showing enhanced to 
 polymer population growth rates.  In that cut, we eliminated systems 
 in which the logarithmic derivative of the total number of polymers with
 respect to Gillespie time was less than 1.

\begin{figure}
 \includegraphics[width=3.5in]{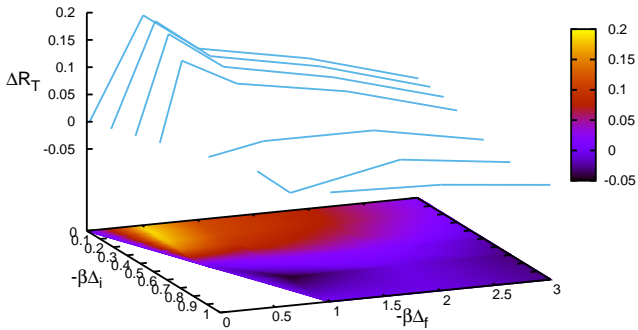}
 \caption{\label{fig:3DmapWcQY11619}Enhancement in $R_T$ arising as a
 consequence of a qeunch as a function of the initial and final 
 values $\beta\Delta_i$ and $\beta\Delta_f$ of the parameter $\beta\Delta$. 
 For all these quenches $p=0.0761$, $b=4$, 
 $l_{max} = 7$ and the quench was applied at reaction step =$10^5$. 
 Each data point is an average over 400 simulations (80 networks and 
 5 realizations per network)}
                \end{figure}

We made a study of the dependence of the observed $R_T$ quench enhancement on 
 the 
 initial and final temperature parameters $\beta\Delta_i$ and
 $\beta\Delta_f$ for the case $b=4$ with 
 results shown in Fig. \ref{fig:3DmapWcQY11619}. High initial temperatures
 ($|\beta\Delta_i|$ small) and relatively high final temperatures ($|\beta\Delta_f|$ also small but larger 
 than $|\beta\Delta_i|$)
 are favored. In the case that $|\beta\Delta_i|$ is 
small though larger than $|\beta\Delta_f|$, there is a wide range of 
$|\beta\Delta_f|$ which is predicted to give substantial $R_T$ enhancement in
the quench
 However, in the envisioned application, final values of 
 $\beta\Delta_f$ are expected to be as large as -4 and in that case,
 the range of $|\beta\Delta_i|$ which gives a large enhancement is
 predicted to be quite narrow. Such estimates can be useful to experimentalists exploring
 the parameter space to determine the conditions under which lifelike
 systems are most likely to be produced by quenching.
 
\begin{figure}
 \includegraphics[width=3.5in]{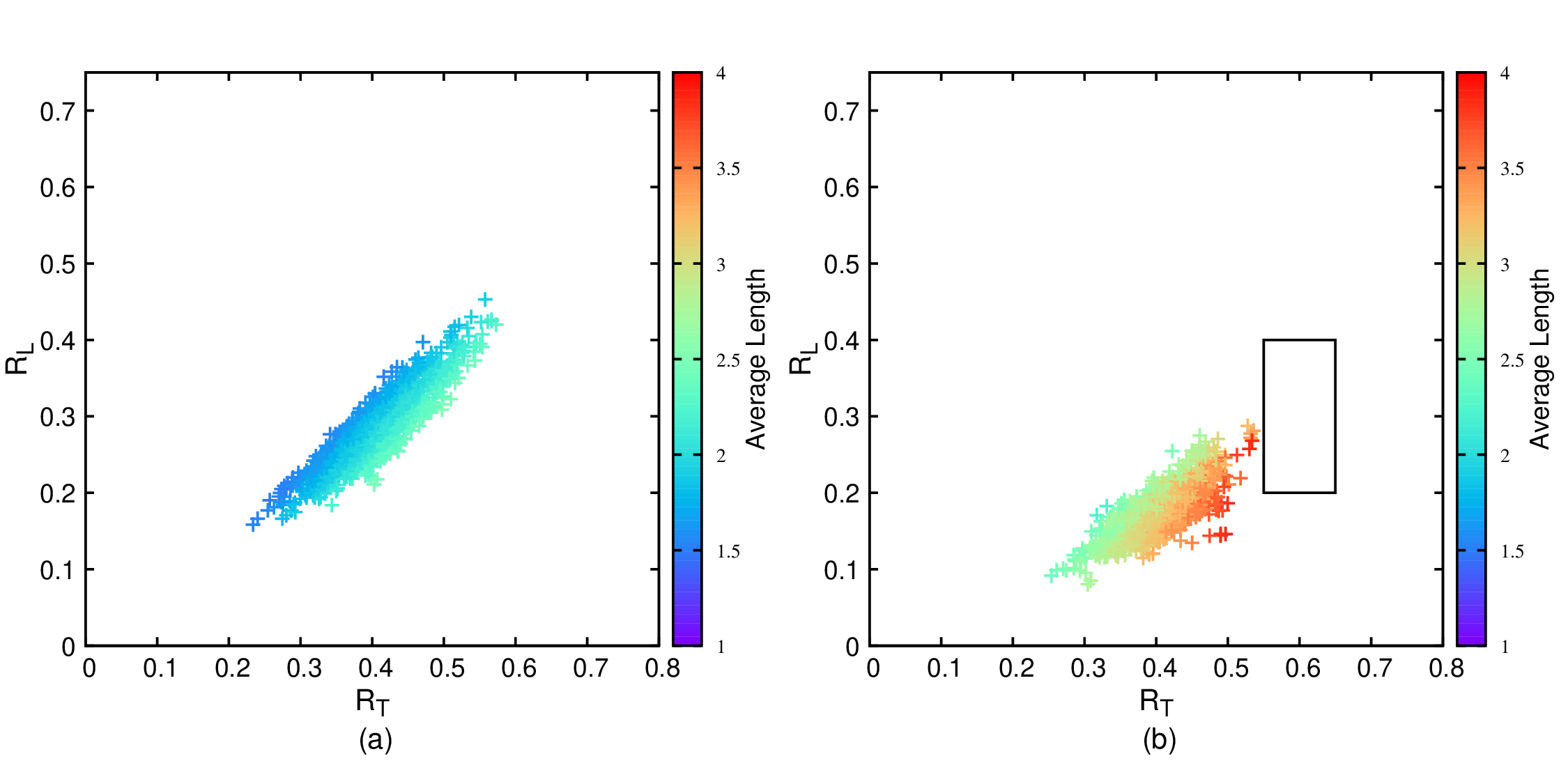}
 \caption{\label{fig:B4L7}At right: Values of $R_L$, $R_T$ found in 
1,200 simulations of the model with $b = 4$, $l_{max} = 7$ and $p = 0.0905$, 
$\beta\Delta_i = -0.01$, $\beta\Delta_f = -4$. At left: Results with
the same parameters and networks but with $\beta\Delta = -4$ throughout the runs (no quench) as a control. The box indicates the region in which real proteomes were found in reference 
 \cite{intoyhalleypopulations}. The color scale indicates the average 
polymer length in the final, quenched state.} 
 \end{figure}

 We explored the distribution of $R_L-R_T$ values in the final simulation
 states in systems running at low temperatures with the corresponding
 distributions when the final state is at the same final temperature
 but has been quenched from a high temperature.  Results are shown in
 Fig. \ref{fig:B4L7} for $b=4$, $l_{max}=7$ and $p=0.0950$ $\beta\Delta_i=-0.01$
 $\beta\Delta_f=-4.$
 Remarkably, more states with longer average polymer lengths appear 
 in the high $R_T$ part of $R_L-R_T$ plane.  These states are quite
 close to the region of the $R_L-R_T$ plane where the values for real proteomes
 are found \cite{intoyhalleypopulations} as indicated by the box in the Fig.
and the results indicate that the quench has stabilize the bonded, long
polymers.

\begin{figure}
\includegraphics[width=3.5in]{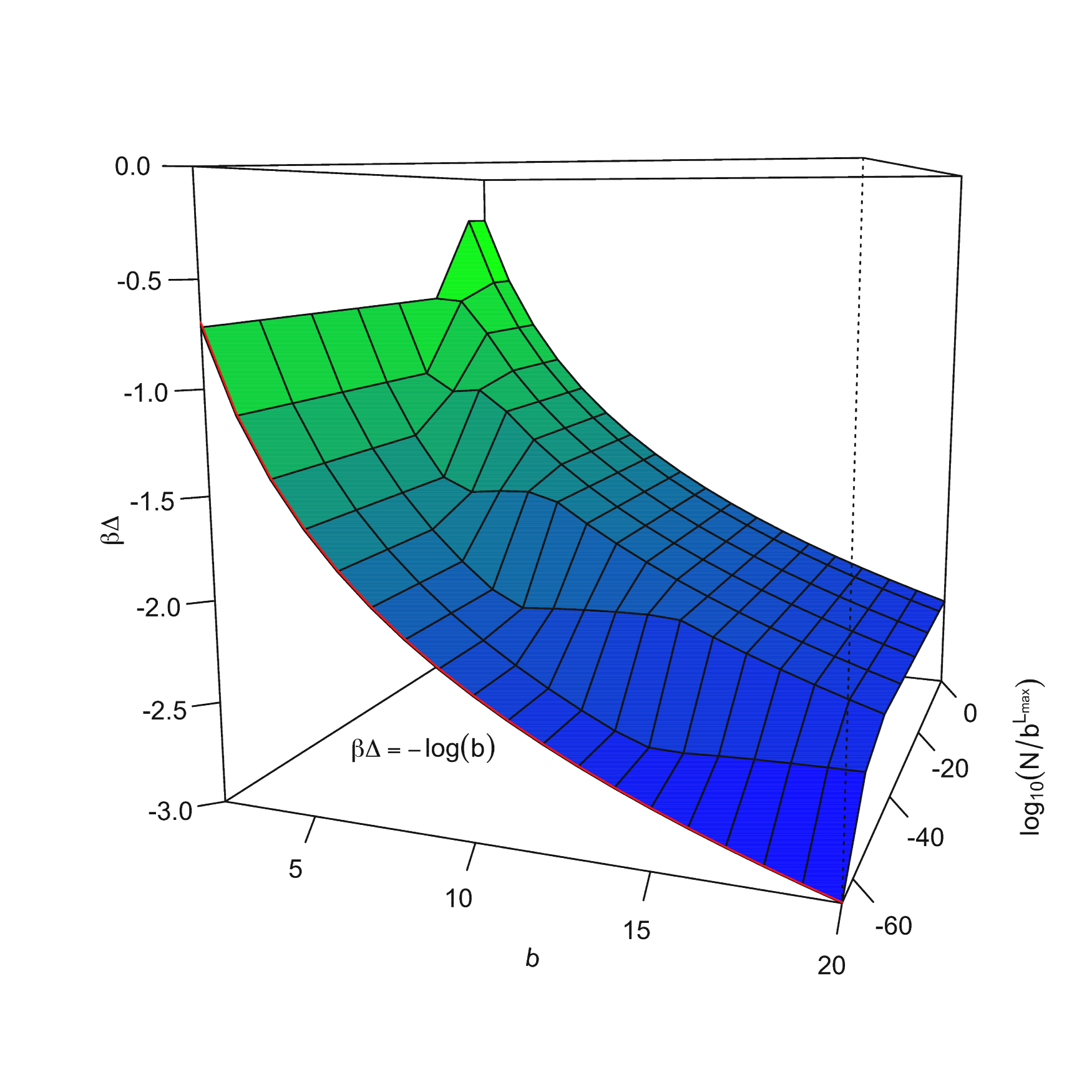}
\caption{\label{fig:hp4fig} $\beta\Delta_{flat}$ as a function of $b$
and $N/b^{l_{max}} $. The red line is the relation $\beta\Delta_{flat} =-\ln b$.}
\end{figure}

Hypothesis 4 is a statement about equilibrium. The values of $\beta\Delta$ 
 at which
$(1/l_{max})\delta \overline{N_L}/\delta L=0 $ are evaluated numerically from
$b=2$ to $b=20$ various values of $N/b^{l_{max}}$ in Fig. \ref{fig:hp4fig} . They
converge to the $\ln b = -\beta\Delta $ when  $N/b^{l_{max}}<< 1$ 
as shown analytically in the Appendix. For proteins in a proteome, 
$N$ is of the order of $10^6$ and $b^{l_{max}} \approx 20^{2000}=4 \times 
10^{2000}$ so the limit is easily realized. 

\section{Discussion and Conclusions} \label{sec:discussion}

In summary, the simulations reported here on a previously developed model
are consistent with
the idea posed in our earlier paper, namely that the polymer length distributions
observed in existing proteomes might suggest that early life was associated with
a higher temperature environment and that the lifelike systems generated in that
environment could have been stabilized by a rapid quench to lower temperatures.
In doing those simulations we introduced new features in the simulations
(but not in the model): Quenches from high to low temperature were incorporated
and the dependence of the resulting low temperature steady states on the initial
temperature, the final temperature, the number of available  monomers $b$ and the 
maximum polymer length was explored. Of particular note is the strong dependence
of the results on the initial, high temperature which needs to be quite precisely
tuned to minimize $R_T$ to achieve maximum enhancement of lifelike states in the final
, low temperature environment. We understand this qualitatively as arising because
the low $R_T$ in the high temperature state permits rapid bond breaking and formation 
allowing a full exploration of the state space by the dynamics.  The sharpness of 
the region associated with the minimum in $R_T$ arises from the abrupt transition,
in the corresponding equilibrium state, 
from an equilibrium  state of nearly all monomers to a state of nearly all maximum length 
polymers as a function of temperature. That was illustrated for proteome 
data from the biosphere in Fig. \ref{fig:RTfuncdb} and is also manifested in 
our simulations. An example from the simulations was shown in Fig. \ref{fig:RTvsdbmodelb} 
( This is closely associated mathematically
with the Bose-Einstein transition in low temperature physics. However we are
working with a finite system, meaning that there are no true phase transitions,
we are ultimately concerned with nonequilibrium states
and the physics is entirely different and not directly associated with any 
quantum effects.) 
Though we are only able to explore it up to $b=7$ we are able to plausibly
 extrapolate to
 the protein relevant value of $b=20$ to confirm that the model
 is approximately consistent with our earlier conjecture 
 \cite{intoyhalleypopulations} that, from the observed length distributions
 in the proteome data, we could infer that the prebiotic formation 
 of the first proteomes formed at a temperature of about $-\Delta/(k_B\ln b)$
The previously proposed 
relation for the optimum initial high temperature (minimizing $R_T$) of $\ln b =
-\beta\Delta $ is approximately consistent with the numerical data extrapolated
to large $b$.

Though we have applied our analysis using the ideas described here to proteome
 data, the same considerations and model  might also apply in principle 
to synthesis of RNA in an RNA-world scenario for the origin of life. The
phosphate bonds in RNA are similarly of higher energy in water than the
separated nucleotides (so in our formulation $\Delta <0$ as for 
proteins ).  Biology only
uses four nucleotides in RNA so we would set $b=4$. The maximum length
of RNA in biological systems is much longer than it is for proteins so
the expansion parameter $N/b^{l_{max}}$ used in the expansion in the 
Appendix is likely to be small.  However experimental production of 
collections of nucleotides without any proteins has proved problematic
in attempts to experimentally produce models for RNA world scenarios and
in contemporary biology RNA generally does not seem to occur without
accompanying proteins though RNA replicases have been
found which could in principle catalyse biochemical reactions
without proteins in an RNA world. On the other hand, relatively isolated systems of proteins 
such as amyloids and prions exist. These features made it difficult to 
find data relevant to the RNA world hypothesis with which to compare
the results of our model and, for those and other reasons, we have not
yet fully explored the possibility of the applicability of our results for 
the $b=4$ case to nucleic
acids in prebiotic chemistry. Note that all of the high temperatures 
associated with optimal conditions for bond formation in the hypotheses
of section \ref{sec:Hyp} will not coincide as closely in our model 
when $b=4$ as 
they do when $b=20$. That would suggest that the high temperature
most favorable to quenching of solutions of 
amino acids to form polypeptides may be more precisely defined than it would
be for formation of RNA from nucleotides.  Thus, at the optimal
high temperature,  our quenching
mechanism might work better for proteins than for RNA. 

We believe that these results may have  implications for possible scenarios for
the origin of life and also for possible laboratory experiments exploring 
conditions which could lead to lifelike chemistry in nonbiological contexts.
With regard to the former we note that in ocean trenches\cite{trenches} liquid
water at temperatures well above the boiling point under ambient conditions is 
continuously being emitted and spilled quite rapidly into cooler water.  
Such encounters of very hot alkaline water emerging from an ocean trench, for
example in a `black smoker', with acidic ocean water at temperatures near 0$^o$ C have 
been suggested\cite{barge} as possible prebiotic sites where electrochemical processes could lead to the 
generation of energy carrying small molecules , particularly FeS, that could  
provide energy for peptide bond formation. In such an environment, we envision that 
formation of  polypeptide networks behaving as in our models might occur and grow.
If such processes continued   from a very early stage in the earth's history
then a very high rate of continuous quenches could proceed over many
hundreds of millions of years.  That a few of those very numerous quenches
could have resulted in trapping of nonequilibrium dynamic systems out of equilibrium
leading in one case to life initiation seems to us at least as plausible as
many alternative scenarios which have been proposed. 
 
An advantage of the scenario discussed 
here is that bonds can be broken and reformed at a high rate in the high temperature
phase, thus allowing a wide exploration of the state space, and then be rendered  more stable  by
quenching.  Most of those low temperature states will not be lifelike, but if this event
occurs many millions of times, some of them may be. 
The function of the quench in the envisioned scenario is that it could trap those states
out of equilibrium with the lower temperature ambient environment associated with the
quenched state, thus possibly permitting a promising lifelike configuration which would be rapidly
transformed in the high temperature state to evolve and grow in the lower temperature quenched
environment. Our simulations are probably underestimating the magnitude of such a trapping
effect, because they do not take explicit account of
the possibility that the quench could stabilize lifelike states because of
the existence of free energy barriers to the hydrolysis reaction leading
to scission of peptide bonds. Such barriers are
 known to lead to survival of some peptide bonds for as long
as centuries in the absence of enzymes\cite{scissionbarrier}, though a
lower limit of more like 35 days is likely. Building a model to take explicit
account of the existence of such barriers is a high priority for future work
and is underway.  In preliminary work in this direction, we are making 
the distribution of 
reaction rates $v$ temperature dependent to take account of 
activation free energies.  
 
A similar quenching phenomenon might
occur in tidal pools, where the daily cycles of drying and wetting are 
accompanied by cooling and heating.  The temperature differences are not
expected to be as large, but an advantage is that the process may be repeated 
many times on the same system. Bond breaking is also sensitive to pH of the aqueous 
environment \cite{pHdependence} and a similar cycling of pH might lead to similar effects in both
the ocean trench and tidal pool contexts.  All these possibilities require 
further theoretical and experimental study.

With regard to laboratory experiments, the experiments of Yin et al \cite{yin} 
in which solutions of amino acids are dried at high temperature and then 
 redissolved in water for analysis approximate some of the
 conditions envisioned here for prebiotic evolution. 
 Matsuno et al \cite{matsunoa}-\cite{matsunoc} did  laboratory experiments in which  solutions of amino
 acid monomers were quenched to low temperature and pressure and length
 enhancements in the polypeptides produced were observed. 
 Our preliminary analysis of the experiments described in
 \cite{yin} and \cite{matsunoa}-\cite{matsunoc} gives low values
 of $R_L$ and much larger values of $R_T$ nearer 1, in qualitative agreement 
 with our simulation results.  However the effects in the
 experiments are larger than in the simulations:  The experimental $R_L$
 values are smaller and the experimental values of $R_T$ are nearer 1 than
 they are in the simulations.  There are several possible
 reasons for the discrepancy including the primitive character of the
 model, effects of unrealistically small $b$ or 
 the  lack of  barriers to
 dissolution of the bonds in the model. Because the 
 experiments of \cite{yin} and \cite{matsunoa}-\cite{matsunoc} share
 a similar quantitative discrepancy with the simulations, it is unlikely
 that the failure to model the details of the drying part of the
 experiment of \cite{yin} is the source of the discrepancy.
A more detailed analysis of these experiments using the measures
employed in this paper will appear later.

\section{Acknowledgements}
Work was supported by NASA grant NNX14AQ05G, by the Minnesota Supercomputing
Institute and by the Open Science Grid.

\section{Appendix: Low density expansion of the equilibrium model}

Following methods closely related to standard derivations\cite{halleySM} of the
virial expansion for gases of interacting atoms, we obtain an 
expansion for $N/b^{l_{max}}$ in the fugacity $z=e^{\mu\beta}$ for the 
total number of particles
in equilibrium as follows: Rewrite equation (\ref{eq:Ncond} ) as 
\begin{equation}
N/b^{l_{max}} =\sum_{L=1}^{l_{max}}{{{(b^{L-l_{max}} -b^{-l_{max}}})z}\over{(e^{\epsilon_l\beta} -z)}}
\label{eq:Nconda}
\end{equation}
where $\epsilon_L=-\Delta(L-1) $. Expand for small $z$ :
\begin{equation}
N/b^{l_{max}} =\sum_{L=1}^{l_{max}}\sum_{n=0}^{\infty} 
z^{n+1}e^{-(n+1)\epsilon_L\beta} (-1)^n(b^L-1)b^{-l_{max}}  
\label{eq:Ncondrearr}
\end{equation}
or reversing the orders of summation
\begin{equation}
N/b^{l_{max}}=\sum_{n=0^{\infty}} z^{n+1} F_n(\beta\Delta)
\end{equation} 
The sums on $L$ in $F_n$ are geometric giving
\begin{equation}
F_n(\beta\Delta) =
\end{equation}
$$
\left ( {{b(1-e^{l_{max}(\ln b +(n+1)\beta\Delta)})}\over{1-e^{\ln b +(n+1)\beta\Delta }}} -
{{1-e^{(n+1)\Delta \beta l_{max }}}\over{1-e^{(n+1)\beta\Delta }}}\right )(1/b^{l_{max}})
$$
To obtain an expansion for $z$ as a function of $\rho=N/b^{l_{max}}$ from
this one inverts order by order in the standard way.  
The $n=0$ term with $b>>1$ gives 
$N/b^{L_{max}} \equiv \rho = ze^{\Delta\beta(l_{max}-1)}$ . 
That is the `Gibbs limit'.
Proceeding similarly for $\delta \overline{N_L}/\delta L $ we take 
$\delta L =1$ and evaluate
\begin{equation}
\delta \overline{N_L}/\delta L  = \left({ {(b^{l_{max}} -1)z}\over
{e^{\Delta\beta(l_{max}-1)}-z}}-
{{(b-1)z}\over {1-z}}\right )=
\end{equation}
$$
		\sum_{n=0}^{\infty} G_n z^{n+1}
$$
with
\begin{equation}
G_n=\left [ e^{-\beta\Delta(n+1)} \left ( e^{l_{max} (\ln b +\beta\Delta (n+1))}-
\right . \right . 
\end{equation}
\begin{equation}
\left .\left .
e^{l_{max}(\beta\Delta (n+1))} \right ) -b+ 1 
\right ]
\end{equation}
Keeping only the n=0 term and setting the result to zero we have
\begin{equation} 
e^{-\beta\Delta} \left (e^{l_{max}(\ln b +\beta\Delta)} -e^{l_{max}\beta\Delta} \right ) -b+1 =0
\end{equation}
with solution for $\beta\Delta_{flat}$
\begin{equation}
\beta\Delta_{flat} =-\left (1/(l_{max}-1)\right ) \ln \left ( {{b^{l_{max}} -1}\over {b-1}}\right )
\end{equation}
giving, when $b>>1$, $\beta\Delta_{flat}=-\ln b$ consistent with hypothesis 4. 

\end{document}